\documentclass{aastex}
\usepackage{spr-astr-addons}

\begin{document}

\title{Discovery of deep eclipses in the cataclysmic variable IPHAS~J051814.33+294113.0}
%% Running heads
\shorttitle{Eclipses in IPHAS~J051814.33+294113.0}
\shortauthors{Kozhevnikov}

\author{V. P. Kozhevnikov}
\affil{Astronomical Observatory, Ural Federal University, Lenin Av. 51, Ekaterinburg 
620083, Russia e-mail: valery.kozhevnikov@urfu.ru}

%\email{\emaila}
\begin{abstract} 

Performing the photometric observations of the cataclysmic variable IPHAS~J051814.33+294113.0, we discovered very deep eclipses. The observations were obtained over 14 nights, had a total duration of 56 hours and covered one year. The large time span, during which we observed the eclipses, allowed us to measure the orbital period in IPHAS~J051814.33+294113.0 with high precision, $P_{\rm orb}=0.20603098\pm0.00000025$~d.  The prominent parts of the eclipses lasted $0.1\pm0.01$ phases or $30\pm3$~min. The depth of the eclipses was variable in the range 1.8--2.9~mag. The average eclipse depth was equal to $2.42\pm0.06$~mag. The prominent parts of the eclipses revealed a smooth and symmetric shape. We derived the eclipse ephemeris, which, according to the precision of the orbital period, has a formal validity time of 500 years. This ephemeris can be useful for future investigations of the long-term period changes. During the latter four observational nights in 2017 January, we observed the sharp brightness decrease of IPHAS~J051814.33+294113.0 by 2.3~mag. This brightness decrease imitated the end of the dwarf nova outburst. However, the long-term light curve of IPHAS~J051814.33+294113.0 obtained in the course of the Catalina Sky Survey during 8 years showed no dwarf nova outbursts. From this we conclude that IPHAS~J051814.33+294113.0 is a novalike variable. Moreover, the sharp brightness decrease, which we observed in IPHAS~J051814.33+294113.0, suggests that this novalike variable belongs to the VY~Scl-subtype. Due to very deep eclipses, IPHAS~J051814.33+294113.0 is suitable to study the accretion disc structure using eclipse mapping techniques. Because this novalike variable has the long orbital period, it is of interest to determine the masses of the stellar components from radial velocity measurements. Then, our precise eclipse ephemeris can be useful to the phasing of spectroscopic data.

\end{abstract}

\section{Introduction}

Cataclysmic variables (CVs) are interacting binary stars consisting of a white dwarf and a late-type companion filling its Roche lobe and transferring material to the white dwarf. In non-magnetic systems, the white dwarf accretes material through an accretion disc. If the white dwarf possesses a strong magnetic field, the transferring material forms a column, and accretion occurs through this column onto one of the magnetic poles of the white dwarf.  Such magnetic CVs are called polars. Intermediate polars are systems, in which a non-synchronously spinning white dwarf accretes material through a truncated accretion disc onto both magnetic poles. Therefore, intermediate polars generate coherent brightness oscillations. CVs have very bright accretion discs and accretion columns, which are much brighter than stellar components. Therefore, in CVs, stellar components are difficult to detect.  Due to accretion processes, CVs exhibit a variety of periodic, quasi-periodic and aperiodic variability phenomena. Depending on the presence of large brightness outbursts, CVs are subdivided into dwarf novae and novalike variables. In dwarf novae, outbursts are thought to occur due to temporary transitions of the accretion disc from a low-temperature state into a high-temperature state.  Due to a large accretion rate, novalike variables reveal permanently hot accretion discs. Eclipsing CVs are important because they make it possible to accurately measure the component masses and to study the accretion disc structure using eclipse mapping techniques. Comprehensive reviews of CVs are given in \citet{ladous94}, \citet{warner95} and \citet{hellier01}.

\citet{witham07} reported the discovery of 11 new CV candidates by the Isaac Newton Telescope Photometric H$\alpha$ Survey (IPHAS). \citeauthor{witham07} performed the ollow-up observations of only three of them. These observations, however, gave insufficiently definite results. Indeed, the spectroscopic observations of two CV candidates (IPHAS~J013031.90+622132.4 and J051814.34+294113.2) gave the ambiguous orbital periods due to aliasing problem (see Fig.~6 in \citeauthor{witham07}). \citeauthor{witham07} discovered that the third CV candidate (IPHAS~J062746.41+014811.3) is an eclipsing CV. This discovery, however, was not all the truth because later \citet{aungwerojwit12} found that IPHAS~J062746.41+014811.3 is an eclipsing intermediate polar. This finding is very important because eclipsing intermediate polars are rare. Moreover, \citeauthor{witham07} highlighted another system, IPHAS~J025827.88+635234.9, which might be a high-luminosity object reminiscent of V~Sge. Performing the photometric observations of this CV candidate, we discovered that IPHAS~J025827.88+635234.9 is an ordinary eclipsing CV \citep{kozhevnikov14}. Although \citeauthor{witham07} published the data of new CV candidates 12 years ago, until recently no comprehensive photometric observations of IPHAS~J051814.34+294113.2 were made. In this paper, we inform of deep eclipses, which we discovered by performing the photometric observations of IPHAS~J051814.34+294113.2 (hereafter J0518).  The four CV candidates highlighted in \citeauthor{witham07} represent a curious case because three of them are eclipsing CVs.

\section{Observations}

\begin{table}[t]
{\small 
\caption{Journal of the observations}
\label{journal}
\begin{tabular}{@{}l c c c}
\hline
\noalign{\smallskip}
Date & BJD$_{\rm TDB}$ start & Length & Out-of-eclipse \\ 
(UT) & (-2,457,000)                   & (h)       & magnitude \\
\noalign{\smallskip}
\hline
\noalign{\smallskip}
2016 Feb. 11  &  430.187539 &   4.2  & 16.0 \\
2016 Dec. 2  & 725.201309  &   6.8 & 16.6 \\
2016 Dec.  8 & 731.349913  &   1.0 & 17.0 \\
2016 Dec. 19 & 742.221852 &   2.8 & 15.6 \\
2016 Dec. 20 & 743.090548 &   6.0 & 15.6 \\
2016 Dec. 22 & 745.079370 &   6.5 & 15.6 \\
2017 Jan. 16     & 770.094247 &   4.4 & 15.9 \\
2017 Jan. 17     & 771.090171 &   1.3 & 16.3 \\
2017 Jan. 18     & 772.236717 &   2.6 & 16.5 \\
2017 Jan. 19     & 773.265303 &   1.6  & 16.8 \\
2017 Jan. 22     &  776.102118 &  0.6 & 17.6 \\
2017 Jan. 24     & 778.101956 &   5.7 & 17.5 \\
2017 Jan. 25     &  779.095064 &   9.9 & 17.9 \\
2017 Jan. 27     &  781.243862 &   2.8 & 17.7 \\
\noalign{\smallskip}
\hline
\end{tabular} }

\end{table}

For photometric odservations, we use a multi-channel pulse-counting photometer with photomultiplier tubes that allows us to make continuous brightness measurements of two stars and the sky background. In this photometer, the relative centring of two stars in the diaphragms is fulfilled by means of rotation of the photometer around the optical axis of the telescope and moving two rectangular prisms attached to the sliding carriage along the optical axis. This photometer makes it possible to maintain the relative centring of the two stars in the diaphragms for a long time. The design of the photometer is described in \citet{kozhevnikoviz}.  Moreover, we use the CCD guiding system, which enables the centring of the two stars in the diaphragms to be maintained automatically during the observational night. Although this photometer provides high-quality photometric data even under unfavourable atmospheric conditions, for a long time we could not observe stars fainter than 15~mag, which are invisible by eye (a 70-cm telescope). Now we can do the relative centring of two stars in the photometer diaphragms, one of which is invisible by eye, by using the coordinates of the invisible star and the computer-controlled step motors of the telescope. Recently we made sure that, using this method, we can obtain high-quality photometric data for stars fainter than 16~mag (e.g., V2069 Cyg, \citealt{kozhevnikov17}).

We performed the tentative photometric observation of J0518 in February 2016 using the multi-channel photometer and the 70-cm telescope at Kourovka observatory, Ural Federal University. Although the first observation night allowed us to clearly detect the eclipse in this star for the first time, the observations were not immediately continued because the right ascension of J0518 is roughly $5^h$. This makes it impossible to observe J0518 in spring. Therefore, the follow-up photometric observations were obtained under better observation conditions in 2016 December and 2017 January. These observations were performed during 13 more nights using the same observation technique. The total duration of observations was 56~h. A journal of the observations is given in Table~\ref{journal}. Note that this table contains BJD$_{\rm TDB}$, which is the  Barycentric Julian Date in the Barycentric Dynamical Time (TDB) standard. TDB is a uniform time and, therefore, is preferable. We calculated BJD$_{\rm TDB}$ using the online-calculator (http://astroutils.astronomy.ohio-state.edu/time/) \citep{eastman10}. To check these calculations, we also used the BARYCEN routine in the 'aitlib' IDL library of the University of T\"{u}bingen (http://astro.uni-tuebingen.de/software/idl/aitlib/). During our observations of J0518 the difference between BJD$_{\rm TDB}$ and BJD$_{\rm UTC}$ was constant. BJD$_{\rm TDB}$ exceeded BJD$_{\rm UTC}$  by 69~s (e.g., \citealt{eastman10}).

\begin{figure}[t]
\includegraphics[width=84mm]{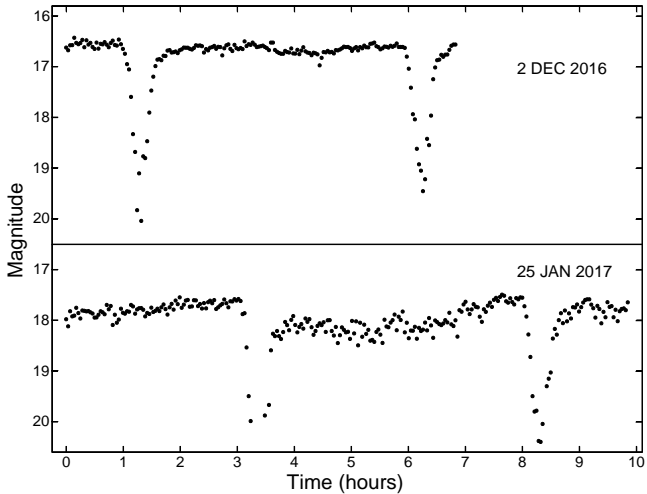}
\caption{Two longest light curves of J0518, which show consecutive eclipses.  Note that the out-of-eclipse magnitude of J0518 is highly variable from night to night} 
\label{figure1}
\end{figure}

The brightness measurements of the programme and comparison stars were obtained using 16-arcsec diaphragms whereas the sky background was measured through a 30-arcsec diaphragm. Data were accumulated with 16-s sampling intervals in white light (approximately 3000--8000~\AA), using a PC-based data-acquisition system. The comparison star is USNO-A2.0 1125-02263093. It has $B=14.6$~mag and $B-R=0.4$~ mag. According to the USNO-A2.0 catalogue, J0518 has the similar colour index, $B-R=0.2$~ mag. This minimizes the effect of differential extinction. During our observations J0518 was much fainter than the comparison star. Therefore, the influence of the variable sky background might be significant.  To diminish this influence, during our observations we periodically removed the stars from the diaphragms and measured the sky background simultaneously in all three diaphragms. The differences of the sky background in the photometer channels were fitted by second- or third-order polynomials, and the sky background was subtracted from the counts of the stars using these fits. Then, we obtained differences of magnitudes of the programme and comparison stars by taking into account the differences in light sensitivity between the star channels.

\section{Analysis and results}

\begin{figure}[t]
\includegraphics[width=84mm]{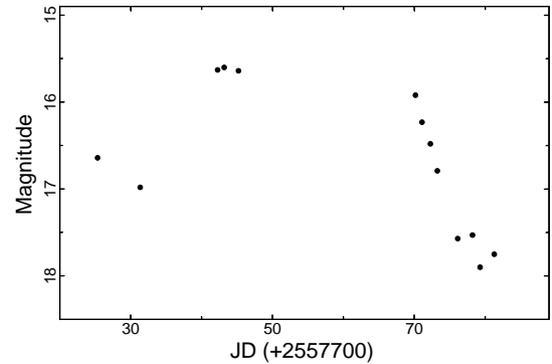}
\caption{Out-of-eclipse magnitudes of J0518 from our observations in 2016 December and 2017 January. Note the sharp brightness decrease by 2.3~mag during the latter four nights}
\label{figure2}
\end{figure}

As mentioned, our photometric observations revealed the eclipses in J0518 for the first time. Fig.~\ref{figure1} presents two longest light curves of  J0518, which show consecutive eclipses. Here and in other appropriate cases, for usability, the differential magnitudes were converted into the magnitudes by adding the magnitude of the comparison star in $B$ band, which seems most appropriate. Although we compare our magnitudes obtained in white light with the magnitude in $B$ band, the systematic errors must not exceed a few tenths of magnitude because the colour indexes of the comparison star and J0518 are close. 

As seen in Fig.~\ref{figure1}, the out-of-eclipse brightness of J0518 greatly varies in the range 16.5--18~mag, and the observed eclipses are very deep. Their depth reaches at least 2~mag. Hence, the deepest points in the eclipses can reach 20~mag. Therefore, to diminish the photon noise, the counts of the stars and the sky background were previously averaged over 128-s time intervals. None the less, some points in the eclipse visible in the lower frame of Fig.~\ref{figure1} on the left are absent because, in these cases, the counts of the sky background  are larger than the counts of J0518. This was not surprising when the star faded up to 20~mag because the sky background was variable. These indefinable points allowed us to roughly evaluate conservative errors caused by the variations of the sky background. The errors can reach ${\pm0.5}$~mag when J0518 approaches  to 19~mag. The errors become less than ${\pm0.2}$~mag when J0518 becomes brighter than 18~mag. These errors monotonically shift large data portions and, therefore, resemble systematic errors.

\begin{table}[t]
%\small
{\scriptsize
\caption{Determination of the period from adjacent eclipses}
\label{table2}
\begin{tabular}{@{}l c c c}
\hline
\noalign{\smallskip}
Date  & BJD$_{\rm TDB}$ ecl. 1 & BJD$_{\rm TDB}$ ecl. 2  & Period \\
(UT)   &  (-2,457,700)                 & (-2,457,700)                      &   (d)     \\
\noalign{\smallskip}  
\hline
\noalign{\smallskip}
2016 Dec. 2    & 25.25612(44) & 25.46220(30) & 0.20608(53) \\ 
2017 Jan. 25   & 79.23670(59) & 79.44217(30) & 0.20547(66) \\
\noalign{\smallskip}
\hline
\noalign{\smallskip}
Average & -- & -- & 0.20578(42)   \\
\noalign{\smallskip}
\hline
\end{tabular} }
\end{table}

Fig.~\ref{figure2} presents the out-of-eclipse magnitudes of J0518 in 2016 December and 2017 January, which were obtained  by averaging of out-of-eclipse light curves. As seen, the out-of-eclipse magnitudes reveal the great changes in the range 15.6--17.9~mag. The out-of-eclipse magnitudes are presented in Table~\ref{journal}. During our observations we obtained ten full eclipses. Because the photometric accuracy was greatly varying depending on the star brightness, we subdivided these eclipses into three groups. When the out-of-eclipse magnitudes were fainter than 17~mag, four eclipses were of low quality because of the absence of several points in some of the eclipses. These eclipses were observed in January~24--27. Two eclipses were of moderate quality due to the noticeable errors of the deepest points. They were obtained in 2016 December~2 when the out-of-eclipse magnitude was 16.6~mag. Four high-quality eclipses were obtained in 2016 February~11, 2016 December~20, 2016 December~22 and 2017 January~16 when the out-of-eclipse magnitudes were brighter than 16~mag and when the deepest points of the eclipses were sufficiently accurate. In addition, in January~18--19 we observed two egresses out of eclipses.

Although two pairs of adjacent eclipses presented in Fig.~\ref{figure1} are of moderate and low quality, none the less, we used these eclipses to find the rough eclipse period, which is apparently equal to the orbital period. The direct measurement of the time span between adjacent eclipses is useful because such a measurement excludes any aliasing problem, which might arise as a result of complicated analysis. The precise mid-eclipse times were found by a Gaussian function fitted to the eclipses, which were cut off at a level of 80 per cent of the eclipse depth. Then, from the time spans between two eclipses we found two orbital periods. The average orbital period is equal to $0.20578\pm0.00042$~d. The results of these measurements are presented in Table~\ref{table2}.

For eclipsing variable stars, the direct fit of an ephemeris to mid-eclipse times gives the best precision of the orbital period. Such a method, however, seems not applicable for our observations of J0518 because the detected eclipses have different quality and because the first eclipse observed in 2016 February 11 is separated by the extremely large gap from the main body of eclipses. Therefore, to increase the precision of the eclipse period, we used the Analysis of Variance (AoV) method \citep{schwarzenberg89}, which is advantageous in comparison with the Fourier transform for non-sinusoidal signals \citep{schwarzenberg98}.  The AoV method usually uses data folded according to a trial period. Here, however, we assign trial frequencies at constant steps of their change and next apply the AoV method. Then, instead of an AoV periodogram, we derive an AoV spectrum, which spans much larger time intervals of variability.

A half-width of the peak at half-maximum (HWHM) in a Fourier power spectrum is often accepted as an error of the period because this conforms to the frequency resolution. However, in most cases such a method gives an overestimated error. \citet{schwarzenberg91} showed that the $1\sigma$ confidence interval of the oscillation period is the width of the peak in the Fourier power spectrum at the $p-N^2$ level, where $p$ is the peak height and $N^2$ is the mean noise power level. Unfortunately, a simple method, which makes it possible to determine the period error in AoV spectra, is unknown. Therefore, we used the above method, which is justified for Fourier power spectra, also for AoV spectra as tentative. 

\begin{figure}[t]
\includegraphics[width=84mm]{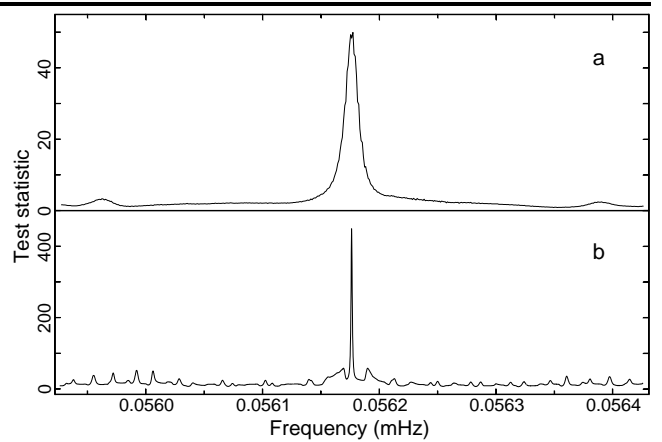}
\caption{Analysis of Variance spectra of J0518 calculated for the data obtained in  2016 December and 2017 January (a) and for four high-quality eclipses obtained between 2016 February 11 and 2017 January 16 (b)}
\label{figure3}
\end{figure}

\begin{table}[t]
{\scriptsize 
\caption{The periods obtained from different methods}
\label{table3}
\begin{tabular}{@{}l c c}
\noalign{\smallskip}
\hline
\noalign{\smallskip}
Method & Period (d) & Dev. ($\sigma$) \\           
\noalign{\smallskip}
\hline
\noalign{\smallskip}
Two pairs of adjacent eclipses              & 0.20578(42)         & 0.6 \\ 
AoV of the December--January data          & 0.2060311(46)     & 0.03 \\
AoV of four high-quality eclipses          & 0.20603098(25)    & -- \\
\noalign{\smallskip}
\hline
\noalign{\smallskip}
\end{tabular}}
\end{table}

\begin{table}[t]
{\scriptsize 
\caption{Verification of the ephemeris}
\label{table4}
\begin{tabular}{@{}l c c c}
\hline
\noalign{\smallskip}
Date     & BJD$_{\rm TDB}$ mid-ecl. & Number    & O--C \\
(UT)     & (-2,457,000)                        & of cycles   &   (s)       \\
\noalign{\smallskip}
\hline
\noalign{\smallskip}
2016 Feb. 11    & 430.219866(96)     & -1519   & $ +0.4\pm10.1   $  \\ 
2016 Dec. 20    & 743.180950(67)     &    0       &  --                           \\
2016 Dec. 22    & 745.241363(70)   & +10        & $ +8.9\pm8.4     $   \\ 
2017 Jan. 16     & 770.170950(98)   & +131      & $ -5.3\pm10.4    $  \\ 
\noalign{\smallskip}
\hline
\noalign{\smallskip}
\end{tabular} }
\end{table}

\begin{table}[t]
{\scriptsize 
\caption{Verification of the period and the error from the pairs of the high-quality eclipses distant from each other}
\label{table5}
\begin{tabular}{@{}l c c c}
\hline
\noalign{\smallskip}
Pair of       & Number    &    Period & Deviation      from \\
eclipses     & of cycles   &   (d)       &  the AoV  period  \\
\noalign{\smallskip}
\hline
\noalign{\smallskip}
Feb. 11 -- Dec. 20   & 1519   & 0.20603100(8)   & $ +0.08\sigma   $  \\ 
Feb. 11 -- Dec. 22   & 1529   & 0.20603106(8)   & $ +0.3\sigma     $   \\
Feb. 11 -- Jan. 16    & 1650   & 0.20603096(8)   & $ -0.08\sigma    $   \\ 
\noalign{\smallskip}
\hline
\noalign{\smallskip}
\end{tabular} }
\end{table}

To calculate the AoV spectrum of J0518, at first we used all individual light curves except for the first light curve obtained in 2016 February 11. This light curve is distant from the main body of light curves, and this might result in aliasing problem. Because J0518 became sometimes very faint, to diminish the photon noise, the counts of the stars and sky background were previously averaged over 128-s time intervals. In addition,   the out-of-eclipse magnitudes were previously subtracted from the individual light curves. The AoV spectrum of the data obtained in 2016 December and 2017 January is presented in the upper frame of Fig.~\ref{figure3}. It shows the distinct peak without noticeable aliases. The precise maximum of the peak was found by a Gaussian function fitted to the upper part of the peak. From this AoV spectrum, we found $P_{\rm orb}=0.2060311\pm0.0000046$~d. Here, the error is found according to \citeauthor{schwarzenberg91}. This error is 5 times less than the HWHM of the peak, which is equal to 0.0000240~d.

Having made certain that the AoV spectrum showed no aliases, we decided to include the distant eclipse obtained in 2016 February~11 into the data analysis to increase the precision of the orbital period.  We, however, discovered that, in this case, the data obtained with a 128-s time resolution are ineligible because the peak in the AoV spectrum calculated at small steps of the frequency change shows a stairs-like structure. Therefore, we excluded the light curves, in which the out-of-eclipse magnitudes were fainter than 16~mag. In addition, we excluded the light curve obtained in December 19, which contained no whole eclipse. Then, we calculated the AoV spectrum using only four light curves containing high-quality eclipses, which were obtained with a time resolution of 16~s. The calculated AoV spectrum is presented in the lower frame of Fig.~\ref{figure3}. As seen, this AoV spectrum also shows no noticeable aliases. From this AoV spectrum, we found $P_{\rm orb}=0.20603098\pm0.00000025$~d. The error is found according to \citeauthor{schwarzenberg91}. This error is 10 times less than the HWHM of the peak, which is equal to 0.00000240~d. The orbital periods obtained by using different methods are presented in Table~\ref{table3}. All these periods are compatible with each other.

We measured the mid-eclipse times of four high-quality eclipses obtained with a time resolution of 16~s by using a Gaussian function fitted to the eclipses, which were cut off at a level of 80 per cent of the eclipse depth. Using the most precise mid-eclipse time and the most precise orbital period we obtained the following ephemeris:
{\small
\begin{equation}
\small BJD_{\rm TDB}(\rm mid-ecl.)=245\,7743.180\,950(67)+0.206\,030\,98(25) {\it E}. 
\label{ephemeris1}
\end{equation} }
Using ephemeris~\ref{ephemeris1}, we calculated the $O - C$ values for the high-quality eclipses and presented them in Table~\ref{table4}. As seen, the $O - C$ values agree with their rms errors. According to the error of the orbital period, the formal validity time of ephemeris~\ref{ephemeris1} is equal to 500 years. Obviously, the large validity time and the high precision of the orbital period are achieved due to the large observational coverage.

\begin{figure*}[t]
\includegraphics[width=174mm]{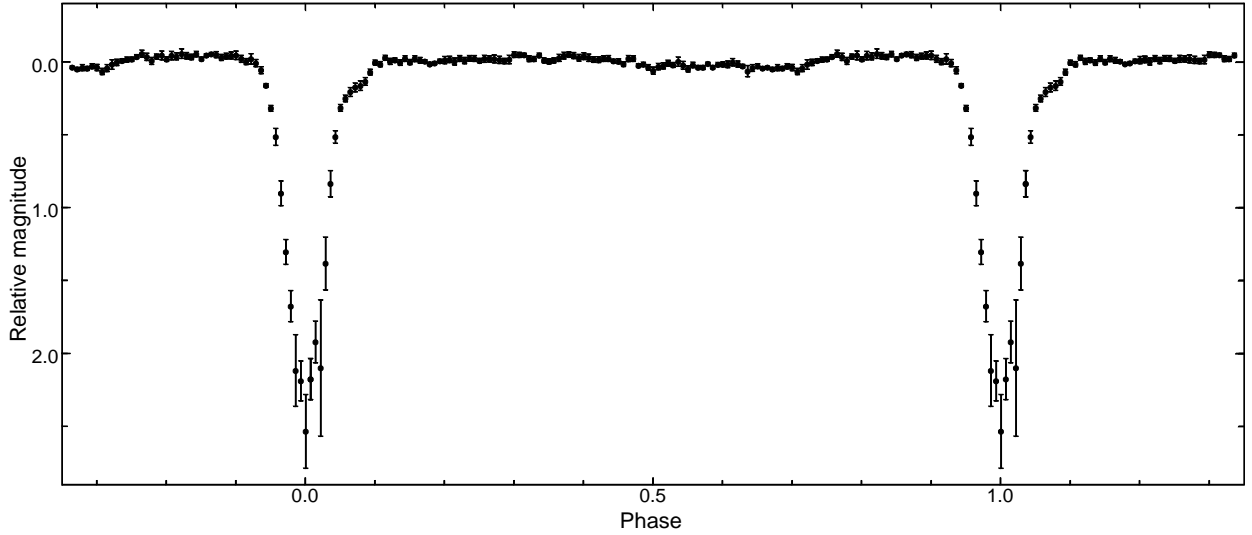}
\caption{Light curves of J0518 folded with a period of 0.20603098~d. We used ten light curves, in which the out-of-eclipse magnitudes were not fainter than 17~mag}
\label{figure4}
\end{figure*}

The large validity time of ephemeris~\ref{ephemeris1} suggests that there is no problem in the determination of numbers of cycles within our observations. Then, we can check the orbital period and its error using the pairs of the high-quality eclipses distant from each other and the time spans between them. The results are shown in Table~\ref{table5}. All obtained periods are consistent with each other. Moreover, these periods are consistent with the orbital period obtained from the AoV spectrum presented in the lower frame of Fig.~\ref{figure3}. The rms errors of the periods found from the rms errors of mid-eclipse times are three times less than the error of the period obtained from the AoV spectrum, which is used in ephemeris~\ref{ephemeris1}. Hence, the error of the period found from the AoV spectrum according to \citeauthor{schwarzenberg91} is overestimated. Therefore, we must regard this error as a conservative error. Then, the estimated validity time of ephemeris~\ref{ephemeris1} should also be conservative. Hence, the real validity time of ephemeris~\ref{ephemeris1}, which corresponds to a $1\sigma$ confidence level, can be a few times large. 

\begin{figure}[hptb]
\includegraphics[width=84mm]{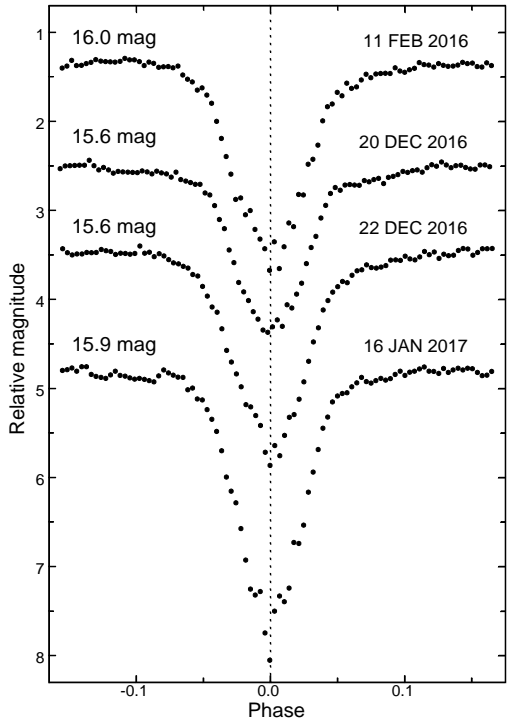}
\caption{Detailed view of four high-quality eclipses, which are obtained with a time resolution of 64~s and placed according to ephemeris~\ref{ephemeris1}. The eclipses are shifted for clarity. For each case, the out-of-eclipse magnitude is shown on the left}
\label{figure5}
\end{figure}

Using the most precise orbital period, we folded ten light curves of J0518, in which the out-of-eclipse magnitudes were not fainter than 17~mag.  The result is shown in Fig.~\ref{figure4}. As seen, before the eclipse, one can notice the small brightness increase, which lasts roughly 0.2 phases and reaches 0.05~mag. This increase is probably caused by the hot spot, which arises in the place where the accretion stream impacts the accretion disc. Moreover, after the prominent part of the eclipse, one can see the appreciable depression, which lasts roughly 0.05 phases and reaches 0.3~mag. This depression can be caused by the asymmetry of the accretion disc. 

From the folded light curve (Fig.~\ref{figure4}), we determined the average eclipse depth by using a Gaussian function fitted to the average eclipse. In contrast with the mid-eclipse times, where we used truncated eclipses, in this case we used the whole eclipse and two adjacent parts of the out-of-eclipse light curve, each of which was of the same length as the length of the whole eclipse including extended eclipse wings. The average eclipse depth is found equal to $2.42\pm0.06$~mag.

Fig.~\ref{figure5} presents the detailed view of four high-quality eclipses. All eclipses including the distant eclipse perfectly coincide in phase according to ephemeris~\ref{ephemeris1}. Inspecting by eye these eclipses, we ticked the phases where the steepness of the eclipse light curves is sharply changing. These phases roughly coincide in all eclipses and are equal to $\pm(0.05\pm0.005)$. Hence, the prominent parts of the eclipses last $0.1\pm0.01$~ phases or $30\pm3$~min. If we include the extended wings of the eclipses, the eclipse duration will be roughly 1~h. The shapes of the prominent parts of the eclipses seem smooth and symmetric.

\begin{figure}[t]
\includegraphics[width=84mm]{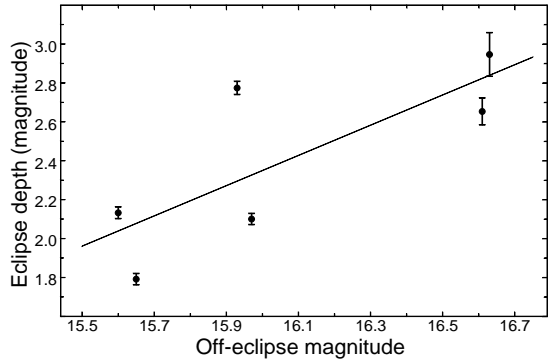}
\caption{Correlation between the eclipse depth and the out-of-eclipse magnitude of J0518}
\label{figure6}
\end{figure}

As seen in Fig.~\ref{figure5}, the depths of the eclipses greatly change. We measured the depths of the high-quality eclipses and the moderate-quality eclipses in the same way how we made this for the average eclipse in the folded light curve and presented these measurements as the function of the out-of-eclipse magnitude (Fig.~\ref{figure6}). As seen, the eclipse depths are varied in the range $1.8-2.9$~mag.  On average, the eclipse depth increases when J0518 becomes fainter. However, the scatter of points in Fig.~\ref{figure6} is large. The slope of the linear fit visible in Fig.~\ref{figure6} is equal to $0.78\pm0.32$. The slope is not statistically significant because this slope exceeds its rms error only by 2.4 times. Therefore, with the small but appreciable probability, the real slope can be equal to 0.  We, however, can definitely conclude that the eclipse depth cannot be noticeably decreasing when J0518 becomes fainter. Obviously, if we suppose such a decrease, the difference between the supposed slope and the observed slope will exceed 3$\sigma$.

\begin{figure}[t]
\includegraphics[width=84mm]{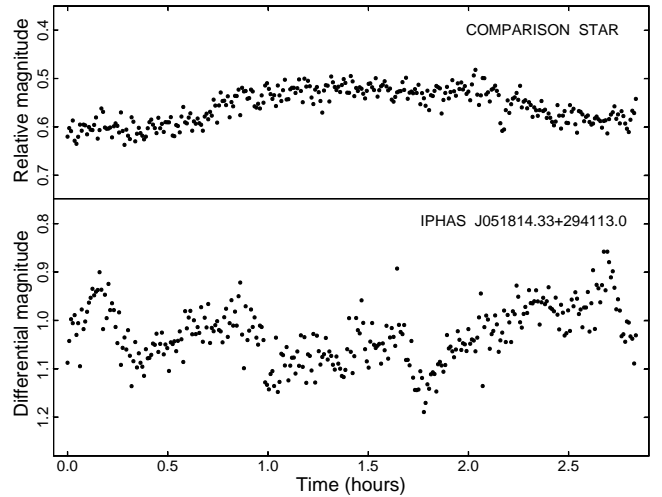}
\caption{Out-of-eclipse light curve of J0518 obtained with a time resolution of 16~s in 2016 December~19. This light curve shows distinct flickering with a peak-to-peak amplitude of about 0.2~mag}
\label{figure7}
\end{figure}

All CVs reveal random brightness changes, which are visible directly in light curves. Such brightness changes are named flickering. The light curves presented in Fig.~\ref{figure1} are rather rough both in vertical scale and in time resolution. Therefore, they show no obvious flickering. We chose the out-of-eclipse light curve obtained with a time resolution of 16~s under good atmospheric conditions in 2016 December~19 when J0518 was sufficiently bright and had a low noise level. This light curve is presented in Fig.~\ref{figure7}. As seen, it shows distinct flickering, where the flickering peak-to-peak amplitude is roughly equal to 0.2~mag. Thus, we have no doubt that J0518 is a CV.

We analysed the out-of-eclipse light curves obtained with a time resolution of 16~s to search for rapid coherent oscillations similar to oscillations in intermediate polars. We limited themselves to nine light curves when the out-of-eclipse magnitudes of J0518 were not fainter than 17~mag. The data of 2016 February~11 were excluded due to aliasing problem for smooth and sinusoidal signals. Our comprehensive Fourier analysis did not reveal rapid coherent oscillations exceeding the noise level. The noise semi-amplitudes were roughly 10~mmag at frequencies higher than 0.7~mHz. At lower frequencies, the noise peaks had much higher semi-amplitudes due to flickering. 

At low frequencies, we attempted to find peaks with periods close to the orbital period. Such peaks might indicate superhumps. We did not find such peaks.   The maximum semi-amplitude of the noise peaks in the vicinity of the orbital frequency was roughly 40~mmag. Hence, superhumps, which exceed 40~mmag in semi-amplitude, are absent.

\section{Discussion}

Performing the photometric observations of the poorly studied CV J0518, we detected the eclipses in this star for the first time. These eclipses allowed us to safely measure the orbital period, which is the most important parameter of any binary star, $P_{\rm orb}=0.20603098\pm0.00000025$~d. The precision of the period is high. This is the result of the large observational coverage, which is equal to one year. As seen in Fig.~\ref{figure3}, we obtained the sufficient observational data to avoid aliasing problem. This, however, was not the case in the first attempt to measure the orbital period of J0518 from spectroscopy made by \citet{witham07}. \citeauthor{witham07} obtained the very different orbital period (see page 1166 in \citeauthor{witham07}). As \citeauthor{witham07} themselves noted, the periodogram presented in their Fig.~6 is heavily aliased, and therefore the obtained orbital period could be incorrect. Thus, the eclipses discovered in J0518 allowed us to correct the orbital period derived earlier from spectroscopy with a great error.

Although during our observations in 2016 December and 2017 January the out-of-eclipse brightness of J0518 revealed great changes in the range 15.6--17.9~mag, which imitate a dwarf nova outburst (Fig.~\ref{figure2}, Table~\ref{journal}), it seems premature to announce that this CV belongs to dwarf novae because in the past this star was frequently observed as sufficiently bright. Indeed, when we observed J0518 for the first time, it revealed an out-of-eclipse magnitude of 16.0~mag. When J0518 was discovered by \citeauthor{witham07}, it was also a relatively bright star of 16.5~mag. In addition, J0518 is listed as a relatively bright star of 15.0~mag in the USNO-A2.0 catalogue. Moreover, the orbital light curve of J0518 is not typical of dwarf novae. Dwarf novae usually reveal pronounced orbital humps caused by bright spots in accretion discs, which have duration of about 0.5 in phase (e.g., \citealt{ladous94}). In contrast, J0518 shows the barely perceptible orbital humps of much lesser duration (Figs.~\ref{figure1} and~\ref{figure4}). Such behaviour is typical of novalike variables (e.g., \citealt{ladous94}). Fortunately, the long-term light curve of J0518 was obtained in the course of the Catalina Sky Survey (CSS) (\citealt{drake09}, http://nunuku.caltech.edu/cgi-bin/getcssconedb\underline{ \,}release\underline{ \,}img.cgi). This light curve can help to define the subtype of J0518.

\begin{figure}[t]
\includegraphics[width=84mm]{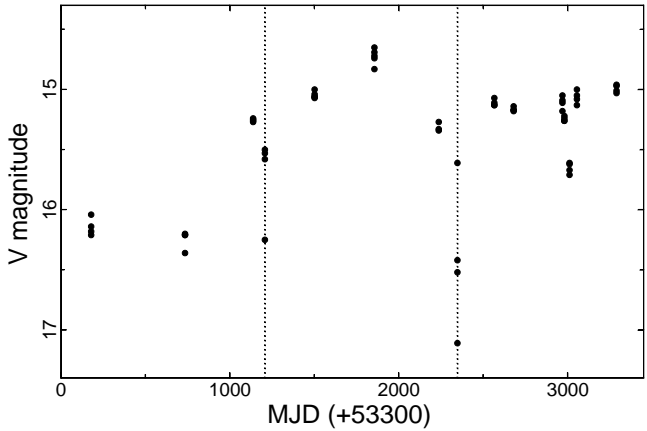}
\caption{Magnitudes of J0518 from the Catalina Sky Survey. The left-hand dotted line marks the case when one of four points belongs to the eclipse. The right-hand dotted line marks the case when three of four points belong to the eclipse}
\label{figure8}
\end{figure}

The long-term light curve of J0518 from the CSS is shown in Fig.~\ref{figure8}. This light curve covers 8 years of observations. It consists of 61 points arranged into groups according to 15 observational nights. Each group consists of 3--5 points covering short time spans in the range 10--50~min. First of all, using this light curve and ephemeris~\ref{ephemeris1}, we can find eclipses. Calculating the $O-C$ values for all points of the CSS light curve, we identified two groups, some points of which lie within eclipses. These groups of points are obtained in 2008 February 12 (MJD 54508.17568--54508.18981) and in 2011 March 28 (MJD 55648.15947--55648.17658). They are marked by dotted lines in Fig.~\ref{figure8}. In the first case, the last of four points has $O-C=-11$~min. and is located at the edge of the eclipse. Three other points have $O-C$ in the range 17--31~min and do not coincide with the prominent part of the eclipse. Because the last point is located at the edge of the eclipse, this point differs from three other points only by 0.7~mag.  The second case is more significant. The third point in the group, which is the faintest point in Fig.~\ref{figure8}, has $O-C=-1$~s and  is located strictly at the mid-eclipse. Two adjacent points have $O-C=\pm8$~min and are located symmetrically at the edges of the eclipse. The first point of the group has $O-C=-16$~min. and is located at the extended wing of the eclipse. This is the obvious reason why the difference of magnitudes between this point and the deepest point is equal to 1.5~mag and is a bit less than the eclipse depths, which we observed in our observations. All other points of the CSS light curve have $O-C$ larger than $\pm20$~min. and do not coincide with prominent parts of eclipses. Therefore, the magnitudes of points in each of these groups differ insignificantly. 

Using the CSS light curve of J0518, we can check the formal validity time of ephemeris~\ref{ephemeris1}. According to the error of the orbital period, the validity time is equal to 500 years.  According to this validity time, the accumulated error of $O-C$ for the faintest point in the group of 2011 March 28 (the right-hand dotted line in Fig.~\ref{figure8}) can be equal to 3.7~min. However, because two adjacent points in this group have nearly equal magnitudes, the real error of $O-C$ must be at least a few times less. If the error was 3.7 min., two adjacent points in this group would differ in magnitude significantly.  Indeed, in this case, one of the two adjacent points should be near the mid-eclipse whereas another point should be shifted from the mid-eclipse by 7 min. Hence, in this case, two adjacent points should reveal noticeably different magnitudes. Thus, we conclude that the error of the orbital period found from the AoV spectrum according to \citeauthor{schwarzenberg91} is a conservative error. Hence, the formal validity time of ephemeris~\ref{ephemeris1} must be noticeably larger than 500 years. This is consistent with the verification of the period and the error from the pairs of the high-quality eclipses distant from each other (Table~\ref{table5}).

As mentioned, the out-of-eclipse light curves of J0518 during 2016 December and 2017 January (Fig.~\ref{figure2}, Table~\ref{journal}) showed changes in the range 15.6--17.9~mag, which resemble a dwarf nova outburst. However, during 8 years of observations the CSS light curve of J0518 (Fig.~\ref{figure8}) showed no strong brightness increases implying dwarf nova outbursts. Except for two eclipses, this light curve showed not very great variations of less than 2~mag in a time scale of years. This behaviour is typical of novalike variables \citep{ladous94}.  Moreover, the CSS light curve showed that, except for two eclipses, there was no deep brightness decrease. Hence, the brightness decrease by 2.3~mag, which we observed at the end of 2017 January (Fig.~\ref{figure2}), is rare and probably short-lived. This brightness decrease resembles low states observed in novalike variables of the VY~Scl-subtype. \citet{honeycutt04} define such low states as a fading of more than 1.5~mag, which lasts less than 150 days. The brightness decrease of J0518 at the end of 2017 January meets this criterion of VY~Scl stars.  Moreover, novalike variables can show stunted outbursts, which have characteristic durations of 5--20 days and amplitudes up to 1~mag \citep{honeycutt04}. We can account for the brightness increase of J0518 happened in 2016 December by such a stunted outburst. Thus, J0518 most probably belongs to novalike variables of the VY~Scl-subtype.

The average eclipse depth in J0518 is equal to $2.42\pm0.06$~mag. The observed eclipses seem very deep. To find out how frequently such deep eclipses occur in other CVs, we examined the catalogue by \citet{ritter03}, the International Variable Star Index database (https://www.aavso.org/vsx) and references therein. We identified 170 eclipsing CVs with known eclipse depths. We learned that most deep eclipses are found in polars. 50 per cent of the eclipsing polars reveal eclipses deeper than 2.4~mag. Polars, however, are systems without discs and have short orbital periods. Therefore, polars seem inappropriate to compare them with J0518. Eclipsing dwarf novae and novalike variables have equal percentages of very deep eclipses, the depths of which exceeds 2.4~mag, namely 13 per cent. Their total number is 20.  Among novalike variables, the deepest eclipse was observed in 2MASS~J22560844+5954299, where the eclipse depth reaches 4.0--4.3~mag in $B$ band \citep{kjurkchieva15}. Among dwarf novae, the deepest eclipse was observed in AR~Cnc, where the eclipse depth is a bit larger than 3~mag in $B$ and $V$ bands \citep{howell90}. In addition, only three CVs with discs have both deeper eclipses and longer orbital periods than the eclipse depth and the orbital period in J0518. These are abovementioned 2MASS~J22560844+5954299 and AR~Cnc and the novalike variable BT~Mon, which has an eclipse depth of about 2.7~mag in white light (see Fig.~1 in \citealt{robinson82}). Thus, our discovery of the deep eclipses in J0518 is a valuable finding because CVs with discs, which have such deep eclipses and such long orbital periods, are rare.

The importance of this finding is strengthened by the fact that, according to the \citeauthor{ritter03} catalogue, only 5 eclipsing CVs are classified as VY~Scl stars. Among them, only two CVs reveal eclipses deeper than 2.4~mag. These are VZ~Scl with an eclipse depth of 2.8~mag in white light \citep{warner75} and GS Pav with a variable eclipse depth of 2.0--3.5 mag in $V$ band \citep{groot98}. Deep eclipses makes it possible to study the accretion disc structure using eclipse mapping techniques (e.g., \citealt{smak98}). For VY~Scl stars, which show no dwarf nova outbursts in low states, such investigations are especially interesting because accretion discs in VY~Scl stars can have peculiarities suppressing dwarf nova outbursts (e.g., \citealt{hameury02}).

Being an eclipsing CV, J0518 is suitable to safely determine masses of its stellar components in future investigations because the eclipses make it possible to determine the orbital inclination. Moreover, J0518 has the long orbital period, and, therefore, its parameters are especially interesting.  As seen in Table 1 in \citet{zorotovic11}, which presents reliable mass determinations in CVs, the most-spread method to determine the stellar masses uses only photometric data. This method can be applied for dwarf novae, in which their complicated eclipse light curves reveal the accretion disc, bright spot and white dwarf eclipses. Then, a physical model of the binary system can give reliable system parameters (e.g., \citealt{littlefair14, wood86}). However, as seen in Table 1 in \citeauthor{zorotovic11}, this method is not appropriate for novalike variables. Obviously, because novalike variables show featureless eclipse light curves, their physical models cannot give reliable system parameters. Hence, to determine the masses of the stellar components in J0518, one should use radial velocity measurements together with photometric data. Such measurements allow obtaining the stellar masses and orbital inclination simultaneously (e.g., \citealt{szkody93, downes86}).  

In CVs, the white dwarf spectrum is directly invisible and, in radial velocity measurements, is replaced by the accretion disc spectrum. Because this spectrum can be contaminated by asymmetric structures of the disc \citep{robinson92}, measurements of radial velocities are difficult. In future spectroscopic observations of J0518, our precise eclipse ephemeris can help to solve this problem due to the phasing of spectroscopic data (e.g., \citealt{hellier01}).

Our precise eclipse ephemeris can be useful for the future investigations of long-term changes of the orbital period in J0518. Such changes might be caused by the changes of the rotational oblateness of the secondary star during its activity cycle \citep{applegate92}. Such changes might also be caused by the possible giant planet orbiting around the centre of gravity in J0518. (e.g., \citealt{bruch14, beuermann11}). The high precision of ephemeris~\ref{ephemeris1}, which is proved by the CSS light curve of J0518, allows calculating the $O-C$ values without ambiguity in the determination of numbers of cycles during centuries. In addition, the large eclipse depth and the small eclipse duration in J0518 make it possible to measure the mid-eclipse times with a precision of 6--8~s (see Table~\ref{table4}). The high precision of mid-eclipse times can facilitate the investigations of the orbital period changes in J0518.

\section{Conclusions}

Performing the photometric observations of the cataclysmic variable J0518, we detected the very deep eclipses in this poorly studied CV for the first time. The comprehensive analysis of our photometric data, which have a total duration of 56~h within 14 nights, yields the following results:

\begin{enumerate}

\item The large time span, during which we observed eclipses, allowed us to measure the orbital period in J0518 with high precision, $P_{\rm orb}=0.20603098\pm0.00000025$~d. 
\item The detected eclipses revealed the variable depths in the range 1.8--2.9~mag. The average eclipse depth was equal to $2.42\pm0.06$~mag. 
\item The prominent parts of the eclipses showed a smooth and symmetric shape. 
\item The prominent parts of the eclipses lasted $0.1\pm0.01$ phases or $30\pm3$~min. 
\item We derived the eclipse ephemeris, which, according to the precision of the orbital period, has a formal validity time of 500 years. This ephemeris is suitable for future investigations of the orbital period changes. 
\item At the end of 2017 January the out-of-eclipse brightness of J0518 has sharply decreased by 2.3~mag. This decrease imitates the end of the dwarf nova outburst. However, the long-term light curve of J0518 obtained in the CSS survey during 8 years revealed no dwarf nova outbursts. Hence, J0518 is most probably belongs to novalike variables of the VY~Scl-subtype.
\item Because J0518 has a long orbital period, this novalike variable is of interest to determine the masses of its stellar components. Because the eclipse shape in J0518 is featureless, the eclipse modelling cannot give reliable component masses.  Hence, to measure the masses, one should use radial velocity measurements. Then, our precise eclipse ephemeris can be useful to the phasing of spectroscopic data.

\end{enumerate}

\section*{Acknowledgments}

This work was supported in part by the Ministry of Education and Science (the basic part of the State assignment, RK No. AAAA-A17-117030310283-7) and by Program 211 of the Government of the Russian Federation (contract No. 02.A03.21.0006). This research has made use of the SIMBAD database, the NASA Astrophysics Data System (ADS), the International Variable Star Index (VSX) database and the Catalina Sky Survey (CSS). The SIMBAD database is operated at CDS, Strasbourg, France. The VSX database is operated at AAVSO, Cambridge, Massachusetts, USA. The CSS survey is funded by the National Aeronautics and Space Administration under Grant No. NNG05GF22G issued through the Science Mission Directorate Near-Earth Objects Observations Program.  This research has also made use of the VizeR catalogue access tool, CDS, Strasbourg, France. The original description of the VizeR service was published in A\&AS 143, 23.

\vspace{1.0cm}
Fig. 1  Two longest light curves of J0518, which show consecutive eclipses.  Note that the out-of-eclipse magnitude of J0518 is highly variable from night to night

\vspace{0.4cm}
Fig. 2  Out-of-eclipse magnitudes of J0518 from our observations in 2016 December and 2017 January. Note the sharp brightness decrease by 2.3~mag during the latter four nights

\vspace{0.4cm}
Fig. 3  Analysis of Variance spectra of J0518 calculated for the data obtained in  2016 December and 2017 January (a) and for four high-quality eclipses obtained between 2016 February 11 and 2017 January 16 (b)

\vspace{0.4cm}
Fig. 4  Light curves of J0518 folded with a period of 0.20603098~d. We used ten light curves, in which the out-of-eclipse magnitudes were not fainter than 17~mag

\vspace{0.4cm}
Fig. 5  Detailed view of four high-quality eclipses, which are obtained with a time resolution of 64~s and placed according to ephemeris~\ref{ephemeris1}. The eclipses are shifted for clarity. For each case, the out-of-eclipse magnitude is shown on the left

\vspace{0.4cm}
Fig. 6  Correlation between the eclipse depth and the out-of-eclipse magnitude of J0518

\vspace{0.4cm}
Fig. 7  Out-of-eclipse light curve of J0518 obtained with a time resolution of 16~s in 2016 December~19. This light curve shows distinct flickering with a peak-to-peak amplitude of about 0.2~mag

\vspace{0.4cm}
Fig. 8  Magnitudes of J0518 from the Catalina Sky Survey. The left-hand dotted line marks the case when one of four points belongs to the eclipse. The right-hand dotted line marks the case when three of four points belong to the eclipse

\end{document}